\documentclass[final,1p,times]{elsarticle} 
\usepackage{graphicx} 
\usepackage{amssymb} 
\usepackage{amsthm} 

\journal{Nuclear Physics A} 
\begin{document} 

\begin{frontmatter} 


\title{Energy Loss and Flavor Dynamics from Single Particle Measurements in PHENIX}

\author{R.~Belmont for the PHENIX Collaboration}

\address{Department of Physics and Astronomy, Station B 1807, Vanderbilt University, Nashville, TN 37235, USA}

\begin{abstract} 

The transverse momentum spectra, yields, and ratios of charged pions, protons, and antiprotons have been studied up to 5 GeV/c in $p_T$ in 5 different centrality classes in Au+Au collisions at $\sqrt{s_{NN}}$ = 200 GeV.  These results are compared and contrasted with the observables calculated in recombination models of hadronization.  They are also used to examine the color charge dependence of parton energy loss in the medium.

\end{abstract} 

\end{frontmatter} 





\section{Introduction}

One of the most intriguing results from the early years of RHIC operation was the apparent non-suppression of baryon production at intermediate values of transverse momentum $p_T$. This result stimulated new models of hadronization mechanisms in the quark-gluon plasma (QGP), many of them based on some form of parton recombination. These models have shown qualitative agreement with the published experimental data, but more detailed results with extended $p_T$ reach are needed to further discriminate between them.

Additionally, the flavor dependence of parton energy loss is expected to become apparent at sufficiently high $p_T$. The difference in energy loss between gluons and quarks propogating a pure color field is the Casimir factor of 9/4, but this effect may be mitigated by parton flavor conversion. By this, we mean elastic scattering processes in which the leading parton of the jet changes flavor or type. The two dominant modes of this are 1) annihilation and its inverse reaction gluon fusion, $\bar{q} + q \leftrightarrow g + g$, and 2) Compton scattering, $q + g \leftrightarrow g + q$, wherein the momenta between the quark and gluon are exchanged and the particle with the higher momentum has consequently changed type~\cite{rainer}.

This study uses the data set collected by PHENIX during RHIC operations in the year 2007, which is over 800 $\mu b^{-1}$ in integrated luminosity of Au+Au collisions at $\sqrt{s_{NN}}$ = 200 GeV. The new results shown herein make use of the high $p_T$ PID capabilities of the time-of-flight and aerogel cherenkov counter detectors in the west arm of the PHENIX spectrometer, which provide track-by-track identification of protons, antiprotons, and charged pions up to a $p_T$ of 5 GeV/c in 5 different centrality classes, 0-10\%, 10-20\%, 20-40\%, 40-60\%, and 60-92\%.  In future studies, higher $p_T$ reach and finer centrality binning will be possible.

\section{Results}


\subsection{Ratios}

\begin{figure}[h]
\begin{center} 
\resizebox{6.5cm}{4.5cm}{\includegraphics{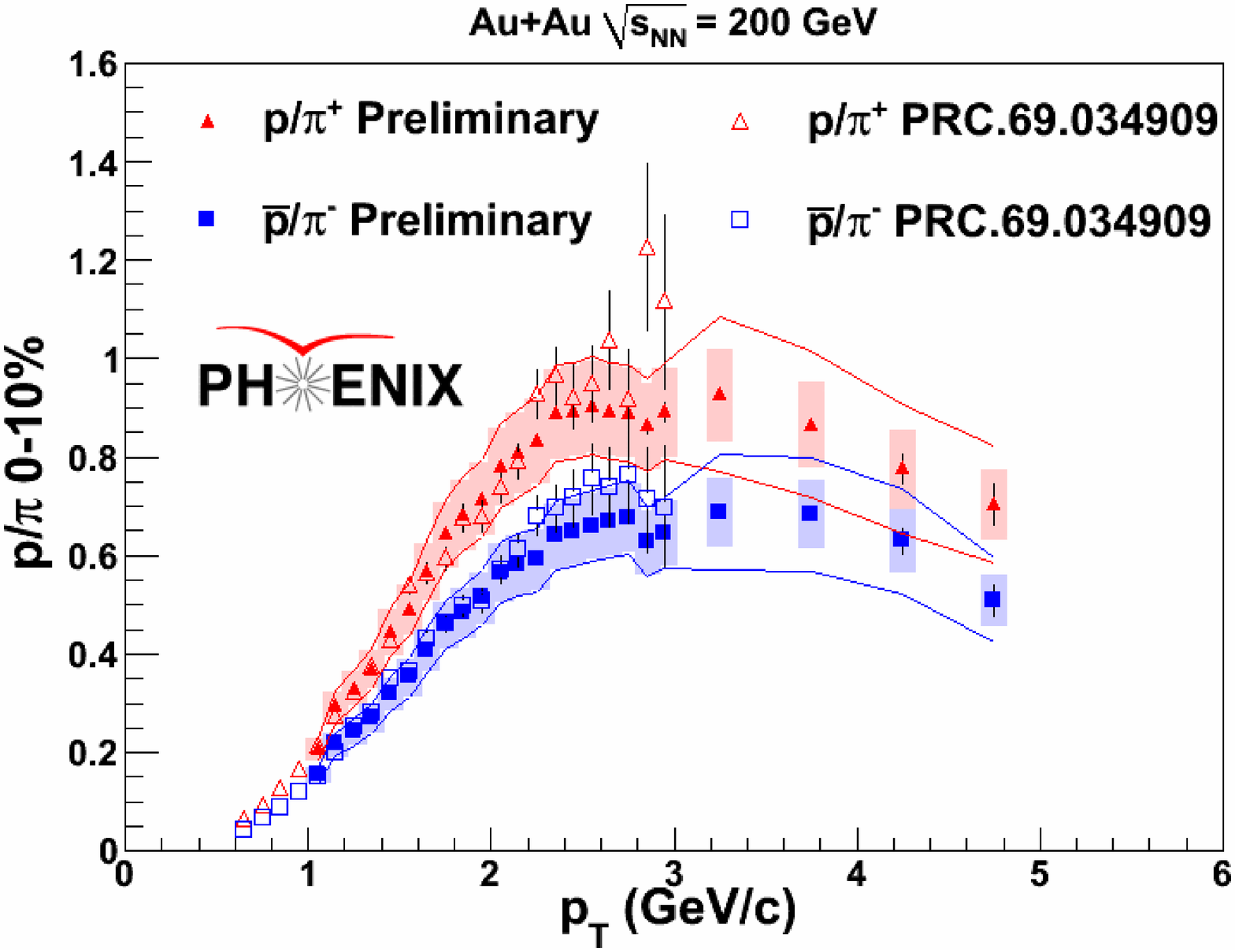}}
\resizebox{6.5cm}{4.5cm}{\includegraphics{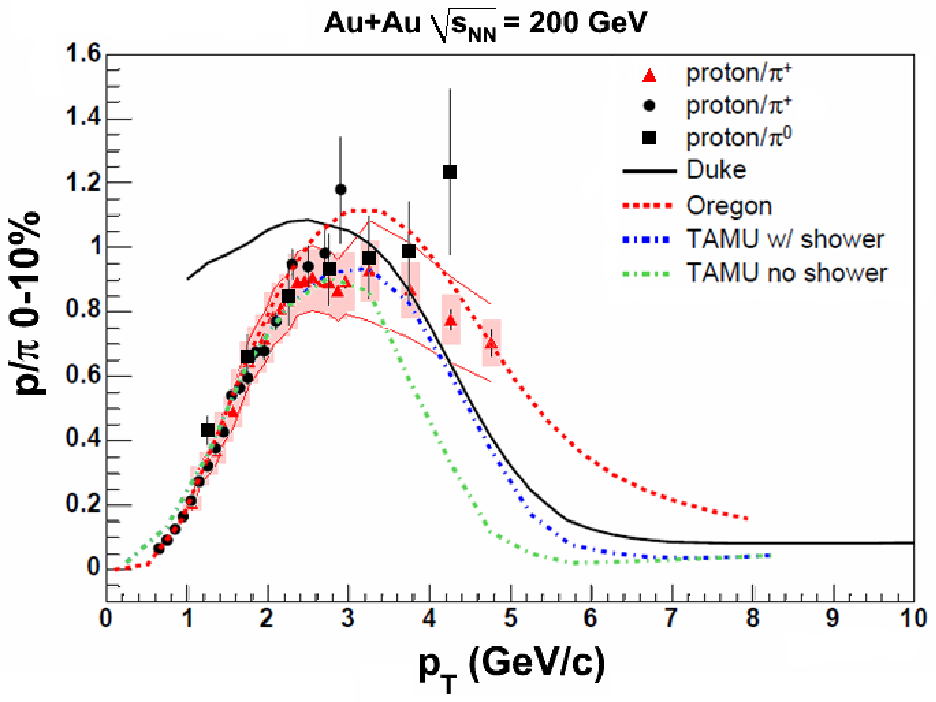}}
\caption{Left Panel: ratios $p/\pi^{+}$ and $\bar{p}/\pi^{-}$ as a function of $p_T$, 0-10\% centrality; data from this study shown in filled circles, data from previous studies~\cite{prl,prc} shown in open squares.  Right Panel: ratio $p/\pi^{+}$ from this study in red circles, ratio $p/\pi^{+}$~\cite{prl,prc} in black circles, and ratio $p/\pi^{0}$~\cite{prl,prc} in black squares, with model comparisons.}
\label{fig:ppi}
\end{center}
\end{figure}

The proton to pion ratio, in both charges, has been examined in each of the 5 centrality classes in this study. Similar to prior observations~\cite{prl,prc}, the $p/\pi^{+}$ and $\bar{p}/\pi^{-}$ ratios in the most peripheral collisions are similar to those in p+p collisions and the maximum of both ratios increases monotonically with centrality. The left panel of Figure~\ref{fig:ppi} shows the new results of the $p/\pi^{+}$ and $\bar{p}/\pi^{-}$ ratios together as a function of $p_T$ for the 10\% most central collisions, and the right panel shows a comparison of $p/\pi^{+}$ only to various hadronization models~\cite{pwp}. The new data, with extended $p_T$ reach, improve the discriminative power against the models. The models adequately describe the data in the low and intermediate $p_T$ regions, but they all fall off too quickly above the maximum, at $p_T \approx 3$ GeV/c.  This disagreement could be from the fragmentation functions used, or from the assumption of the relative contribution of recombination and fragmentation to the final hadron spectrum.

\subsection{Nuclear Modification Factors and Double Ratios}

\begin{figure}[h]
\begin{center} 
\resizebox{6.5cm}{4.5cm}{\includegraphics{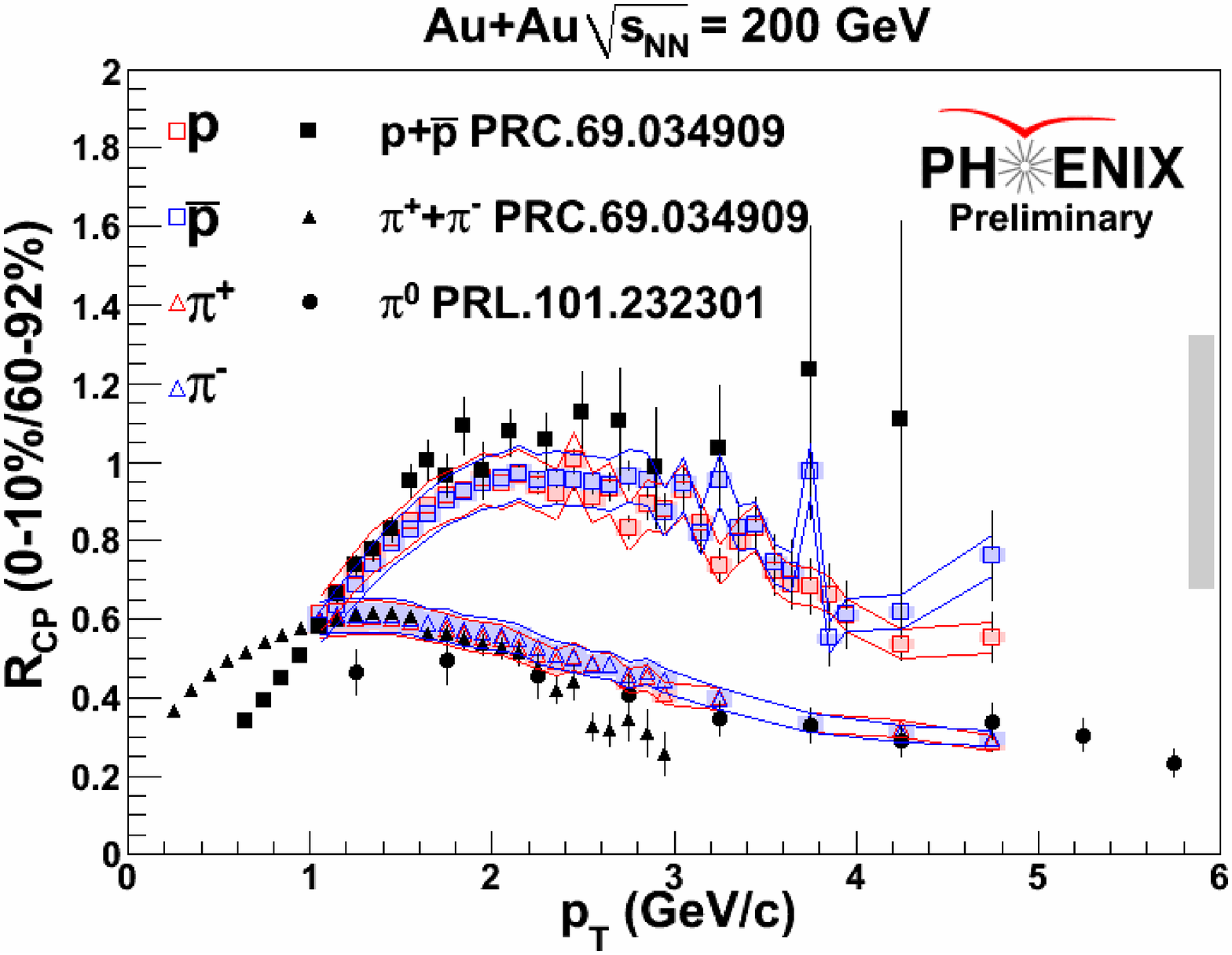}}
\resizebox{6.5cm}{4.5cm}{\includegraphics{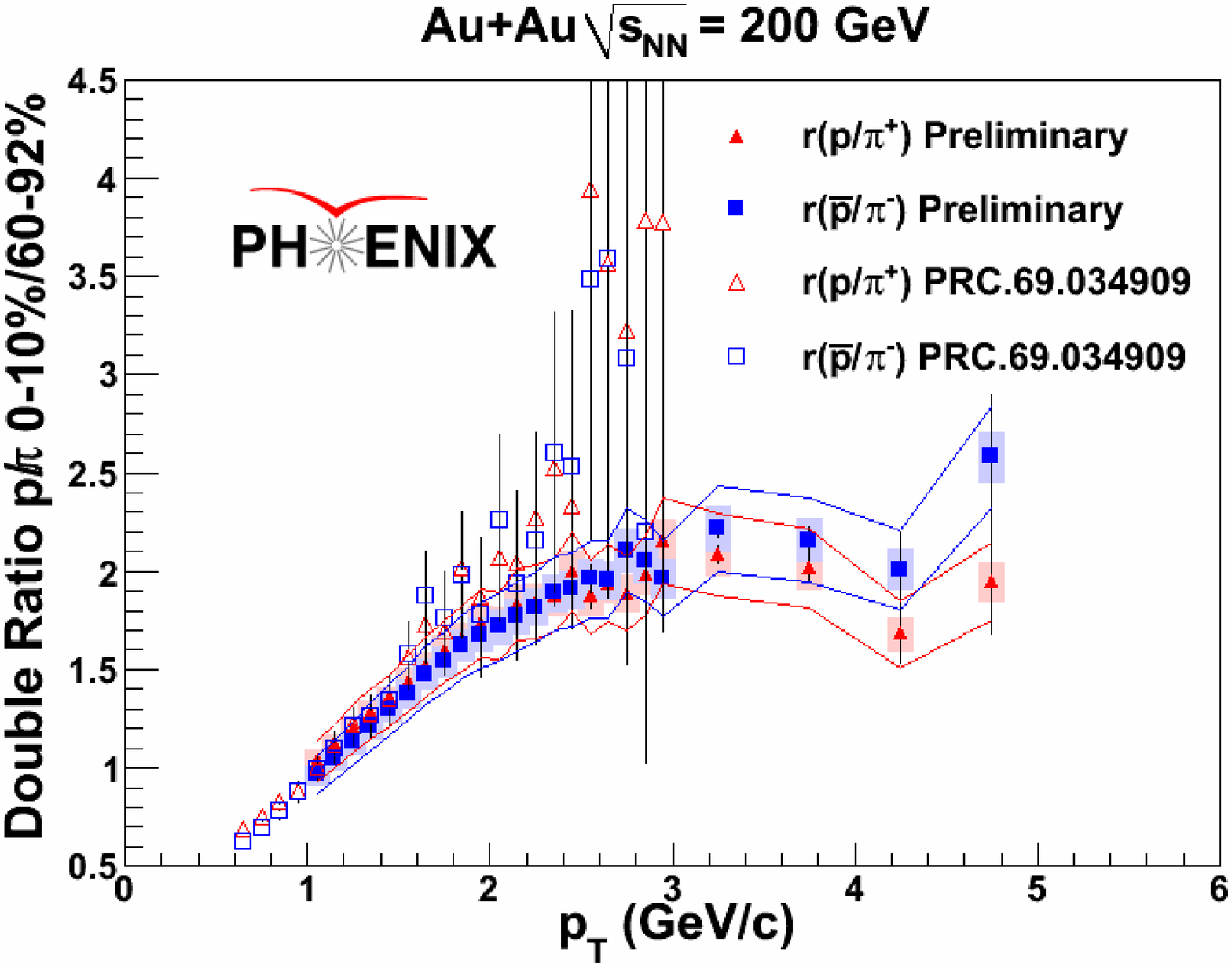}}
\caption{$R_{CP}$ vs $p_T$ of pions and protons (left panel) and double ratio $R_{CP}(p)/R_{CP}(\pi)$ vs $p_T$ (right panel).}
\label{fig:NMF}
\end{center}
\end{figure}

To study the energy loss of hard-scattered partons, we examine the nuclear modification by comparing the hadron yields in central and peripheral collisions scaled by the number of binary nucleon-nucleon collisions, $R_{CP}$. 
The large data sample collected allows for the study of $p$, $\bar{p}$, $\pi^{+}$, and $\pi^{-}$ separately and with significantly improved statistical precision and higher $p_T$ reach compared to previous results~\cite{prl,prc}.
We see in the left panel of Figure~\ref{fig:NMF} that both of the charged pions have have a similar suppression pattern at all $p_T$ and that at higher momentum they agree quite nicely with the neutral pions as well.  Additionally, and perhaps more interesting, the $p$ and $\bar{p}$ also agree within errors at all $p_T$, indicating that the jets that form them have experienced similar suppression. It is expected that antibaryons will have a larger gluon fraction than baryons which, owing to the stronger energy loss expected for gluons, should result in a stronger suppression pattern for antibaryons than baryons.

In Reference~\cite{rainer}, a novel measurement, the so-called ``double ratio,'' is proposed as an observable to test for jet flavor conversion. The fragmentation functions of protons, indeed all baryons, have larger gluon contributions than those of pions, or mesons in general. Considering the expected stronger suppression of gluons, protons should exhibit stronger suppression than pions, meaning the $R_{AA}$ or $R_{CP}$ for protons should be lower than that of pions, and so the ratio $r(p/\pi) = R_{AA}(p)/R_{AA}(\pi)$ should be less than unity.  As seen in the right panel of Figure~\ref{fig:NMF}, the ratio in the available $p_T$ regime is much greater than one, so there is clearly still some interplay between parton coalescence and jet fragmentation.  Note that, at this time, p+p reference data are not available with the appropriate $p_T$ reach, so we need to use $R_{CP}$ as a proxy to $R_{AA}$. In the $p_T$ regions already studied, the $p/\pi$ ratio is comparable but slightly greater than that in p+p, so the conclusion would not change.

\subsection{Comparison with $v_2$}

\begin{figure}[h]
\begin{center} 
\resizebox{6.5cm}{4.5cm}{\includegraphics{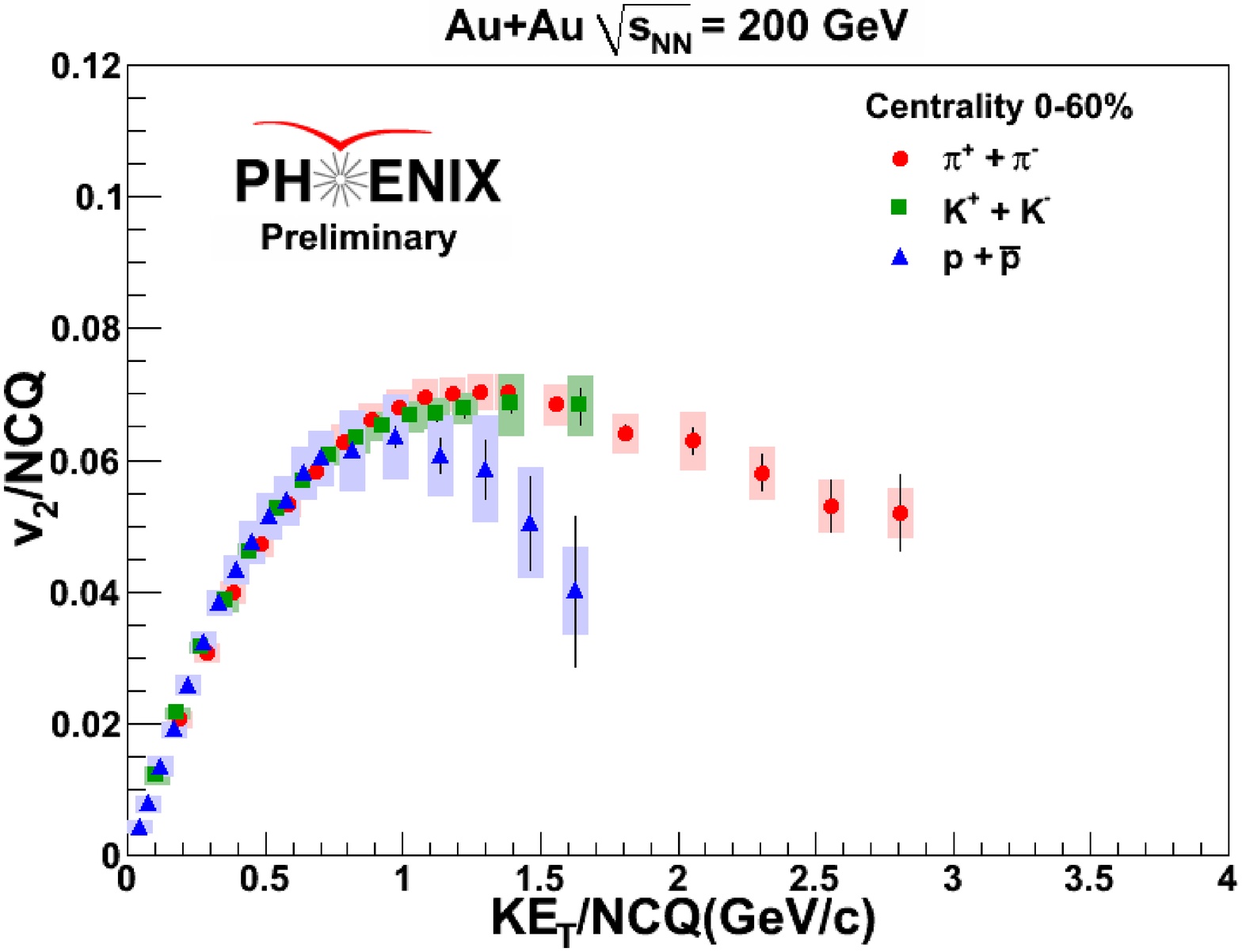}}
\resizebox{6.5cm}{4.5cm}{\includegraphics{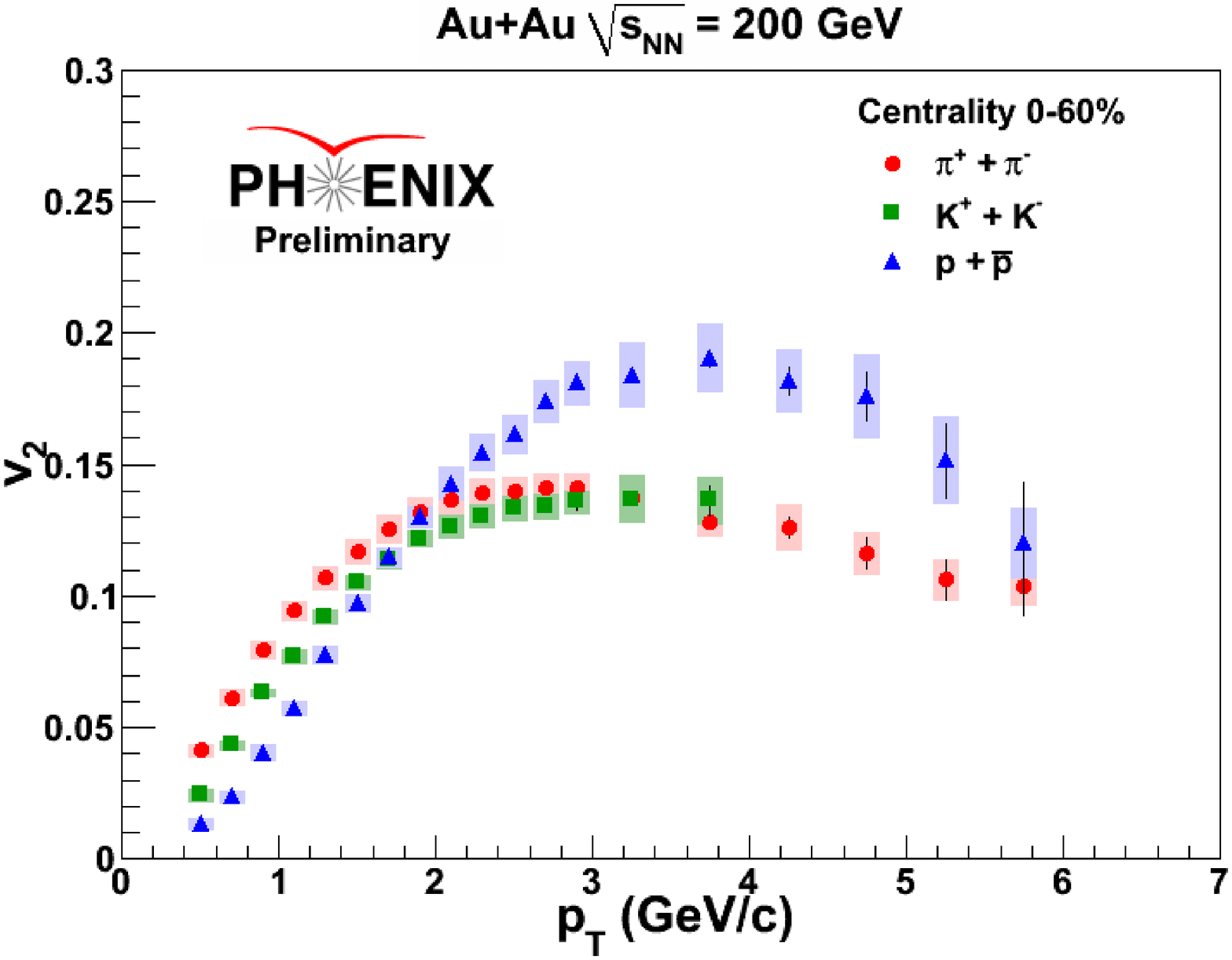}}
\caption{Left panel: anisotropy parameter $v_2$ of identified hadrons as a function of $KE_T$, each axis scaled by the number of constituent quarks~\cite{shengli}.  Right panel: anisotropy parameter $v_2$ of identified hadrons as a function of $p_T$~\cite{shengli}.}
\label{fig:v2}
\end{center}
\end{figure}

Azimuthal anisotropy measurements of identified hadrons can also help us to further understand the hadronization mechanisms.
Anisotropy flow parameters have been studied using the same experimental apparatus as used in this study~\cite{shengli}.
So-called ``constituent quark scaling'' has been known for some time and is a feature of quark coalescence models~\cite{NCQ}.
Previous measurements have focused on low to intermediate $p_T$ where we believe that elliptic flow is the relevant mechanism for the generation of $v_2$.
Elliptic flow is of course a soft physics phenomenon, so at sufficiently high $p_T$ we can expect jet properties to dominate $v_2$.
The left panel of Figure~\ref{fig:v2} shows anisotropy parameter $v_2$ as a function of transverse kinetic energy $KE_T$ for pions, kaons, and protons, with each axis scaled by the number of constituent quarks.
In non-central events, there is less material in-plane and more material out-of-plane, so making the very reasonable assumption that there is some path dependence of the energy loss one naturally comes to the conclusion that $v_2$ should be positive even at high $p_T$, which is seen in the data.  Additionally, the color charge dependence of partonic energy loss should cause larger $v_2$ for gluon jets than for quark jets, and this should manifest as protons having higher $v_2$ than pions. The right panel of Figure~\ref{fig:v2} shows $v_2$ as a function of $p_T$ for pions, kaons, and protons, and the axes are \textit{not} scaled.  In the intermediate $p_T$ region where recombination dominates, the proton $v_2$ is higher than the pion $v_2$, as expected.  However, as one goes to higher $p_T$, where jet fragmentation begins to dominate particle production, we see that the proton $v_2$ starts to decrease and appears to merge with the pion $v_2$. 

\section{Conclusion}

We have herein reported the salient features of our recent study of $p_T$ spectra of charged pions, protons, and antiprotons in Au+Au collisions at $\sqrt{s_{NN}}$ = 200 GeV.  The $p/\pi$ ratio as a function of $p_T$ shows qualitative agreement with the various recombination models, but a more detailed theoretical study with updated fragmentation functions is needed.  On the experimental side, higher $p_T$ reach and reduced systematic uncertainty should be achievable.  The $R_{CP}$ and $v_2$ suggests a strong amount of interplay between recombination and jet fragmentation and flavor dependent effects. 




\end{document}